\documentclass[12pt,preprint]{aastex}

\newcommand{\etal}{{\it et al.}}

\newcommand{\lgpmin}{$\log(P)_{m}$}
\newcommand{\lgpmax}{$\log(P)_{M}$}
\newcommand{\kms}{km s$^{-1}$}

\slugcomment{}

\shorttitle{CEMP-s Binary Fraction}
\shortauthors{Lucatello et al.}

\begin{document}


\title{The Binary Frequency Among Carbon-Enhanced, {\it s}-Process 
Rich, Metal-Poor Stars  \altaffilmark{1, 2}}


\author{Sara Lucatello\altaffilmark{2,3}, Stelios Tsangarides\altaffilmark{4},
Timothy C. Beers\altaffilmark{5}, Eugenio Carretta \altaffilmark{2,6},  
 Raffaele G. Gratton \altaffilmark{2},
Sean G. Ryan\altaffilmark{4} }

\altaffiltext{1}{Based in part on observations collected at the European
Southern Observatory,  Paranal,  Chile (ESO Programme 167.D-0173).}
\altaffiltext{2}{INAF, Osservatorio Astronomico di Padova,  Vicolo dell'Osservatorio 5, 
        35122,  Padova,  Italy.}
\altaffiltext{3}{Dipartimento di Astronomia,  Universit\`a di Padova,  Vicolo dell'Osservatorio 2, 
        35122,  Padova,  Italy.}
\altaffiltext{4}{The Open University, Milton Keynes, MK7 6AA, UK}
\altaffiltext{5}{Department of Physics \& Astronomy and JINA: Joint Institute
for Nuclear Astrophysics,  Michigan State University, 
East Lansing,  Michigan 48824-1116.}
\altaffiltext{6}{INAF, Osservatorio Astronomico di Bologna, via Ranzani 1, 40127, Bologna, Italy} 


\begin{abstract}

We discuss radial velocities for a sample of carbon-enhanced, {\it
s}-process rich,  very metal-poor stars (CEMP-s hereafter), analyzed with high-resolution
spectroscopy obtained over multiple epochs. We find that $\sim$68\% of the stars
in the sample show evidence of radial velocity variations. The expected
detection fraction for these stars, adopting the measured binary fraction in the
field ($\sim$60\%), and assuming that they share the same period and
eccentricity distribution, is $\sim$22\%. Even if one assumes that the true
binary fraction of these stars is 100\%, the expected detection percentage is
$\sim$36\%. These values indicate that the binary fraction among CEMP-s stars is
higher than the field binary fraction, suggesting that {\it all} of these
objects are in double (or multiple) systems. The fact that the observed
frequency of velocity variation exceeds the expected detection fraction in the
case of an assumed binary fraction of 100\% is likely due to a more restricted
distribution of orbital periods for these objects, as compared to normal field
binaries. Our results indicate that CEMP-s stars are the metal-poor analogues of
classical CH-stars.

\end{abstract}

\keywords{stars: AGB and post-AGB --- stars: carbon --- stars: chemically peculiar --- binaries: spectroscopic}
\section{Introduction}

The most extensive spectroscopic surveys undertaken to date to identify large
samples of very metal-poor stars, the HK survey (Beers {\it et al.} 1992; Beers
1999) and the Hamburg/ESO survey (HES hereafter; Christlieb {\it et al.} 2001a,
b; Christlieb 2003), have shown that carbon-enhanced, metal-poor (CEMP hereafter) stars
(here taken to mean [C/Fe] $>$ 1.0) account for up to $\sim$25\% of stars with metallicities lower
than [Fe/H]$\sim -$2.5 \footnote{[A/B]=$\log
\frac{N_A}{N_B}-\log(\frac{N_A}{N_B})_{\odot}$}. Despite extensive
investigations of this class of objects by means of high-resolution spectroscopy
({\it e.g.}, Norris, Ryan \& Beers 1997a, b; Aoki {\it et al.} 2001, 2002a, b, c), the
origin of carbon in these stars still remains unclear. The carbon enhancement
phenomenon appears in stars that exhibit (at least) five different abundance
patterns. A handful of CEMP stars have been identified with {\it no}
enhancements in their {\it n}-capture elements (hereafter, CEMP-no; Norris, Ryan
\& Beers 1997b; Aoki \etal{} 2002a), at least two of which (CS~29498-043; Aoki et al. 2002d,e
and CS~22949-037; Depagne et al. 2002) also appear to exhibit large enhancements
in N, O, and the $\alpha$-elements, while a single case of a highly {\it
r}-process-enhanced CEMP star, CS~22892-052 (Sneden {\it et al.} 2003, and
references therein) has been noted (hereafter, CEMP-r). There also exist several
objects,
 which,
together with very pronounced {\it s}-process enrichment, exhibit
an overabundance in Eu with respect to the {\it s}-process models predictions, 
so that they have been claimed to have been enriched by both 
the {\it r} and {\it s}-process (see {\it e.g.} Cohen \etal 2003).  By far the
most numerous class is represented by CEMP stars characterized by {\it
s}-process-element enrichments (hereafter, CEMP-s). Several CEMP-s stars have
been studied with high-resolution, high signal-to-noise spectroscopy ({\it
e.g.}, Aoki \etal{} 2001, 2002a, b; Johnson \& Bolte 2002; Lucatello \etal{} 2003;
Sivarani \etal{} 2004). Aoki \etal~(2003) recently found, based on a sample of
33 CEMP stars, that over 70\% of their objects with [Fe/H] $< -2.5$ are
characterized by {\it s}-process-element enrichment.

The differences between these five classes suggest that the mechanisms
reponsible for the carbon enrichment in these objects might well be associated
with different astrophysical scenarios. While the number of CEMP-no and CEMP-r 
stars is still small, making proposed enrichment scenarios difficult to
explicitly test, the number of well-studied CEMP-s stars ($\sim$30 so far)
provides a reasonable sample for a statistical analysis of their binary status,
which we undertake in this paper. 

One of the scenarios proposed to explain CEMP-s stars is that they are the
formed by a mechanism that is analogous to that invoked for the origin of the Ba
II, the classical CH, and the subgiant CH stars. The Ba II stars have low
velocities and high metallicities (see {\it e.g.} Jorissen \etal 1998),
 while the classical CH giants exhibit high
velocities and are metal-poor (see {\it e.g.} Bond 1974); 
these giants are not sufficiently luminous
to be AGB stars. Both classes exibit enhancements in C and in {\it s}-process
elements. 
Subgiant CH stars, discovered by Bond (1974), are characterized by a
similarly peculiar abundance pattern and are thought to be progenitors of
moderately metal-deficient Ba II stars (Luck \& Bond 1991). Systematic
spectroscopic studies have shown that essentially all of these stars are members
of binary systems (see, {\it e.g.}, McClure \& Woodsworth 1990). Hence, the
scenario invoked to account for the observed chemical peculiarities in stars of
these evolutionary states is that of accretion of material synthesized by a more
massive intermediate-mass companion star during its AGB phase. Such mass transer
can take place via Roche-lobe overflow (more likely in objects with shorter
periods; see Han \etal{} 1995) or wind accretion (longer periods). The detection
of the expected white dwarf companion stars (see, {\it e.g.}, B\"ohm-Vitense
1980, Dominy \& Lambert 1983, B\"ohm-Vitense \& Johnson 1985) have provided
further support to this scenario.

While there has been speculation that CEMP-s stars might be the metal-poor
equivalent of the classical CH stars (Preston \& Sneden 2001; Sneden, Preston \&
Cowan 2003), conclusive evidence in support of this hypothesis has not yet been
presented. Recent theoretical results, {\it e.g.}, Fujimoto, Ikeda \& Iben (2000)
and Schlattl \etal{} (2002), suggest that low-mass, extremely metal-poor stars
evolve into carbon stars along paths that are quite different from those
followed by more metal-rich stars of younger populations. Carbon may well be
produced through an additional (different) mechanism at low metallicity, {\it
e.g.}, extra mixing at the onset of He-flash. Hence, in order to understand the
formation mechanism of CEMP-s stars it is crucial to first establish whether or
not these objects are all members of binary systems.

\section{Sample Definition and Observations}

We first set a few criteria for the selection of CEMP-s stars. The aim of these
criteria is to clearly characterize the sample and differentiate it from the
classical Ba II, CH, and sgCH stars. Thus, for our analysis we select stars with
temperatures higher than 4800\,K and surface gravities $\log g \geq 1.3$, in order to
rule out likely cases of self-enrichment, which may apply to AGB stars (in any
case, intermediate-mass, metal-poor AGB stars are not expected to be present in
the Galactic halo, due to their comparatively rapid evolution). Moreover, we set
a metallicity upper limit of [Fe/H] $=-$1.8, in order to distinguish CEMP-s stars
from the classical CH-stars, whose typical metallicities extend as low as
$-$1.0 to $-$1.5 (see, {\it e.g.}, Vanture 1992). Thus, our objects have a
metallicity, [Fe/H], which is {\it at least} a factor of two less than that of
classical CH-stars. This separation is set to limit the sample to a metallicity
range for which, as discussed, C-production might possibly occur through different
mechanisms. We also set a lower limit on C-enhancement of 1\,dex ([C/Fe]$\geq
+$1.0). Among the stars that meet both criteria, we have selected those objects
with clear evidence of {\it s}-process enrichment. The atmospheric parameters of
the selected sample stars are listed in Table \ref{t_star_at}.

We have collected new observations for nine CEMP-s stars. The observations for
two objects, CS~22956-028 and CS~29497-034, were obtained using UVES at the
VLT/Unit 2 (Kueyen). The resolving power of these spectra is
$R=\lambda/\Delta\lambda\simeq50,000$, and the spectral coverage ranges from 3600
to 4800\,{\AA} and from 5700 to 9500\,\AA, respectively, in the blue and red
arms; the slit width was fixed at 1 arcsec. The extraction and reduction was
performed using the standard UVES pipeline. 

The remaining seven objects, CS~22880-074, CS~22898-027, CS~29526-110,
CS~30301-015, HD~196944, LP~625-44 and LP~706-7, were observed as part of a
larger programme to monitor the radial velocities of candidate and confirmed
CEMP stars, and calculate abundance patterns for the former (Tsangarides 2004).
The observations are described in detail in that document and are only briefly
summarised here. Three high-resolution echelle spectrographs were used in six
observing runs for this programme: the (now-decommissioned) Utrecht Echelle
Spectrograph (UES: Unger \textit{et al}. 1993) of the William Herschel Telescope
(WHT), Spettrografo ad Alta Risoluzione Galileo (SARG: Gratton \etal 2002)
of Telescopio Nazionale Galileo (TNG) and University College London (Coud\'e)
Echelle Spectrograph (UCLES: Walker \& Diego 1985; Stathakis \textit{et al}.
2000) of the Anglo-Australian Telescope (AAT). The spectra taken with these
telescopes had resolving powers of $R=\lambda/\Delta\lambda\simeq52,000$, 57,000
and 40,000 respectively. We set the slit width to 1.1'' (UES), 0.8'' (SARG) and
1.5'' (UCLES). The raw frames for the seven objects were reduced in
IRAF\footnote{IRAF is distributed by the National Optical Astronomy
Observatories, which is operated by the Association of Universities for Research
in Astronomy, Inc., under cooperative agreement with the United States' National
Science Foundation.}, using standard data reduction procedures. The reduced
spectra cover 3550-5860 \AA\ (UES), 3900-5140 \AA\ (SARG), and 3750-4900 \AA\
(UCLES).

\section{Observed Binary Frequency Among CEMP-s Stars\label{s_bin_of}}

The radial velocities (hereafter, V$_{r}$'s) for the UVES spectra were measured 
using a scheme based on the cross-correlation technique (Tonry \& Davis 1979),
which was developed to measure radial and rotational velocities for globular
cluster dwarfs and subgiant stars; the typical error for these measurements are
$\sim$0.2\,\kms{} for well-exposed spectra (Lucatello \& Gratton 2003).

A detailed description of the procedure to measure the V$_{r}$'s of the
remaining seven sample stars is given in Tsangarides (2004). Here we provide a
brief outline. A cross-correlation technique was adopted for these objects
as well; this uses the metal-poor ($-2.60 \leq [Fe/H] \leq -2.40$; Lambert \&
Allende-Prieto 2002, Nissen \textit{et al}. 2002, Aoki \textit{et al}. 2002),
carbon-mildly enhanced (0.22 $\leq$ [C/Fe] $\leq$ 0.6; Tomkin \textit{et al}. 1992;
Norris, Ryan \& Beers 1997) subgiant HD~140283 as the template. A spectrum of
this object with signal-to-noise ratio greater than 80/1 was obtained for each of the runs
during which the programme stars were observed, and was reduced according to the
same data-reduction procedures as the target spectra.

The geocentric radial velocity of HD~140283 provided the zero-point for the
heliocentric radial velocities we obtained for the seven objects. It was
calculated from the observed spectra, as opposed to being adopted from the
literature, by measuring the shift of several hundred metallic lines, which were
carefully selected discarding possible blends, clearly asymmetric lines, and very
strong lines. The internal error of this calculation is given by dividing the
standard deviation of individual lines with the number of lines used.

Finally, the heliocentric radial velocities of the spectra of the seven program
stars were measured by cross-correlating them with the spectrum of HD~140283
using the IRAF task \textit{fxcor}, then adding the appropriate heliocentric
correction. This procedure produces a second internal error for each object's
heliocentric radial velocity: the deviation of individual pairs of
cross-correlated echelle orders. The errors reported in Table
\ref{t_obs_log} are equivalent to the quadrature sum of the
internal errors, but do not take into account any systematic
effects in the calculated velocity of HD~140283.

The heliocentric radial velocities measured for HD~140283 from each run range
between $-$171.33 $\pm$ 0.08 km s$^{-1}$ to $-$170.63 $\pm$ 0.04 km s$^{-1}$. Latham
\textit{et al}. (2002) monitored HD~140283 for 3114 days with 19 observations,
reported a mean radial velocity of $-$171.12 $\pm$ 0.29 km s$^{-1}$ for this object.
The mean of our determinations, -170.98 $\pm$ 0.22 km s$^{-1}$ is in 
close agreement with the radial velocity reported by Latham {\it et al}. Thus, we
estimate the systematic error affecting the radial velocity of the template to
be 0.30 \kms.

Table \ref{t_obs_log} lists the observation log, the measured V$_{r}$'s, and
their estimated errors. Adding published data in the literature to our sample,
we obtain a total of 19 CEMP-s stars with high resolution, high signal-to-noise
spectroscopic analysis and multi-epoch observations, with a minimum baseline of
$\sim$200 days. The sources of the literature data used in our sample are listed
in Column 4 of Table \ref{t_bin_po}. Orbital
solutions have been derived for several objects in our sample
(see Columns 6 and 7 of Table \ref{t_bin_po}). On
the basis of these data, we calculate the $\chi^{2}$ value for the radial
velocity distribution: 
\begin{equation}
\chi^2=\sum_{i=1}^{n} (\frac{v_{i}-\bar{v}}{\sigma_{v_i}})^2
\end{equation}
for each of the stars
in the sample. Then, we compute the probability, P($\chi^{2}|f$), that the
V$_{r}$ values obtained for the same stars are compatible with different
measurements of the same values, {\it i.e.,} the probability that the observed
scatter is due to observational errors, not to the intrinsic variation of
the measured physical quantity.   

Preston \& Sneden (2001) found that velocity errors derived from multiple
observations of radial-velocity-constant giant stars are larger than the
standard deviations for individual spectra by a factor of $\sim$2-3. Some red
giants are known to exhibit velocity ``jitter,'' (Carney \etal{} 2003), however
this phenomenon appears to affect only stars within $\sim$1\,mag of the red
giant tip, which, adopting an isochrone of metallicity of $-$2.3 and an age
of 12\,Gyr (Yi, Kim, \& Demarque 2003), corresponds approximately to a $\log g
\sim$1.1. Our adopted limit on surface gravity of $\log g\geq$1.3 should
exclude such objects. We choose to multiply the $\sigma$'s derived from our
measurements (as well as those from the literature) by a factor of three, to
allow for systematic errors when comparing radial velocities from different
sources. The factor of three is arbitrary, and acts to reduce the number of
binary detections, therefore it is a conservative choice. It should be kept in
mind that multiplying the measurement errors by this value will distort the $\chi^2$
statistics toward higher values of P for radial-velocity-constant stars. 

The typical errors quoted in the literature for the adopted V$_{r}$'s are
$\sim$1\,\kms. Table \ref{t_bin_po} lists the calculated values of $\chi^{2}$,
degrees of freedom $f$ ({\it i.e.} $f=n-1$, where $n$ is the number of
observations), and P($\chi^{2}|f$) for each one of the 19 CEMP-s stars. The
quantity Q($\chi^{2}|f$)=1$-$P($\chi^{2}|f$), {\it i.e.}, the probability of the
measurement scatter arising from instrinsic variation of radial velocity, is
also listed.

Inspection of Table \ref{t_bin_po} shows that most of the stars in this sample
have a very high probability of being radial-velocity variables. The stars with
derived orbital solutions are consistently found to have very low values of P
(high values of Q), supporting the validity of our test. For such stars we
list the derived orbital elements, along with their source, in Table
\ref{t_bin_orb}. We consider the stars with positive identification of
radial-velocity variability to be those with P$<$0.01 (Q$\geq$0.99). Adopting
this definition, the fraction of stars showing V$_{r}$ variation in our sample
is 68$\pm$11\%. The error has been computed using a binomial distribution.
This value is {\it not} compatible with the most recent estimates of the
spectroscopic binary frequency among normal field stars, such as found by
Carney \etal~(2003) for local metal-poor dwarfs and giants. In fact, these
authors performed an analysis very similar to that used in this work, relying on
the $\chi^{2}$ statistic to discriminate V$_{r}$-variable stars from those with
constant V$_{r}$, and found that a fraction of $\sim$17\% of stars exhibited
detectable V$_{r}$ variations. We stress that the comparison with the results
obtained by Carney \etal{} is meaningful with the underlying assumption that the
binary fraction among field stars is not dependent on metallicity. Strictly
speaking, the observed CEMP-s binary frequency should be compared to that
of C and {\it s}-process {\it normal} stars of similarly low metallicity.
The binary fraction among stars at such metallicity is still not well known.
However, for the high-metallicity end of our sample ([Fe/H] $\sim
-2.0$), it is quite similar to that found for stars of solar metal abundance
(see, {\it e.g.}, Carney \etal{} 2003; Zapatero-Osorio \& Martin 2004). We will
henceforth assume that the binary fraction is independent of metallicity, and
thus adopt for the present discussion a value of $\sim$60\% (Jahrei{\ss} \& Wielen 2000).

We must warn the reader of the potential bias that might affect our sample. When
a star is suspected to be a V$_{r}$ variable, the data might be published faster
than that of an analogous star that does not exhibit variation. This
would introduce a bias in favor of short period objects, and, in principle, in
favor of binaries versus non-binaries. This effect cannot be estimated
quantitatively. However, in most cases the binarity of the objects in our sample
could not be established on the basis of the data from a single author.
Moreover, recently published results constitute only a small fraction of the
dataset. Therefore, this bias is expected to have negliglible impact on the final
results of the present work. 

We emphasize that the sample for this analysis 
was selected {\it exclusively} on the basis of metallicity, C-enhancement,
evolutionary status and observational baseline
without any {\it a priori} knowledge of their showing radial 
velocity variations and/or being known binaries.

\section{Simulations \label{ss_bin_sim}}

It is interesting to compare our reported results with the percentage of the
expected detectable binary stars, for a given binary fraction, which could be
identified as such by our observational scheme. This is accomplished using a
Monte-Carlo simulation. We extract 10,000 datasets, each of which is randomly
assigned to be either a V$_{r}$-constant or a V$_{r}$-variable star according to
the input binary fraction. 

For each of the binary stars, the orbital parameters are assigned randomly
according to appropriate distributions. The orbital inclination, {\it i}, the
longitude at the ascending node, $\omega$, and the initial phase, $\nu_{0}$,
have no preferred values, hence for each we assume a uniform distribution over
the physically meaningful range of values, {\it i.e.}, [0,$\frac{\pi}{2}$],
[0,2$\pi$], and [0,2$\pi$], respectively.

McClure \& Woodsworth (1990) pointed out that the orbital solutions obtained for Ba II and
classical CH stars indicate that their orbital eccentricity, $e$, is typically
lower than those systems containing C- and {\it s}-process elements 
{\it normal} stars, likely because of
the mass transfer which has taken place in such objects. The adoption of an
eccentricity distribution peaked at low values only marginally affects the
simulations, slightly {\it decreasing} the detection probability. Our choice of
a uniform distribution for $e$, within the permitted range [0,1), is thus a {\it
conservative} one.

For the orbital periods, $P$, we adopted the observed distribution by Duquennoy
\& Mayor (1991), which has been measured for field stars and has been adopted in
order to compare them to CEMP-s stars. This distribution is characterized by
$\overline{\log(P) }=$4.8 and $\sigma_{\log(P)}=$2.3, where $P$ is expressed in
days. Given the fact that the longest observation baseline for our sample stars
is only $\sim$12\,years, allowing very long periods, such as could arise from
the use of the full distribution (which peaks at $\sim$120\,years), would only
contribute to the noise. Therefore, we set an upper limit to the period
distribution by discarding those values of $P$ for which the expected orbital
amplitude falls to one-third of the adopted velocity error ($\sim$0.3\,km/s).

The value of the orbital semi-major axis, $a$, is fixed by the extracted period
and the values set for the masses. We assume for the stars under analysis
M$_{1}=$1.0\, M$_{\odot}$ for the mass of the already evolved member of the
pair, now likely a white dwarf, and M$_{2}=$0.8\, M$_{\odot}$ for the mass of
the observed (surviving) star. The choice of such a large value for M$_{1}$
provides conservative estimates of the likely detectable fraction of binaries.
In fact, the use of such a mass, instead of $\sim 0.6$\,M$_{\odot}$, which is
probably more likely, results in a higher probability of radial velocity
variation detection. Moreover, it must be kept in mind that, while the choices
for the values of the masses are reasonable, they are somewhat arbitrary.
Fortunately, they do not considerably affect the value of $a$. The semi-major
axis is proportional to the cube root of the sum of the masses, therefore any
choice of a pair of values within reasonable limits for the stars under analysis
would have a small effect on the derived parameters. 

For each of the simulated stars, either binary or single, we randomly select one
of the observation patterns, $k$ (1$\leq k\leq$19), {\it i.e.,} one of the 19
combinations of the $j_{k}$ time intervals and measurement errors that was
actually used for the $k$th star. For each of the simulated stars, on the basis
of the orbital parameters, and for each one of the time intervals in the
selected observational pattern, we calculate the expected values of V$_{r}$, to
which we added an ``observational error''. The latter is determined as a value
randomly extracted on the basis of a normal distribution whose $\sigma$ is the
observational error attributed to the actual observation. Then, the values of
$\chi^{2}$, P($\chi^{2}|f$) and Q($\chi^{2}|f$) for these simulated observations
are calculated. For consideration of these simulations, we also take P$<$0.01 as
a positive detection of V$_{r}$ variation, the same criteria used for our sample
of real stars. Table \ref{t_bin_proc} shows the result of this procedure,
listing the percentage of the total number of stars detected as V$_{r}$
variables with the described algorithm on the basis of the set criteria and as a
function of the input binary fraction.  

As seen in Table \ref{t_bin_proc}, the percentage of V$_{r}$ variables expected to be
detected by our observational scheme, adopting, as discussed, the measured binary fraction 
in the field ($\sim$60\%, Jahreiss
\& Wielen 2000) is $\sim$22\%. This value is 
somewhat larger than the spectroscopic binary frequency measured by Carney
\etal~(2003), $\sim$17\% for metal-poor field stars.
 Nevertheless, the agreement is reasonable. The small difference
with the Carney \etal~value could be due to our assumption about the masses, as
noted above. Moreover, it should be kept in mind that the simulations were
performed on the basis of the observation patterns for our specific sample,
which are different from the observational patterns of Carney \etal~(2003).

The observed fraction of radial-velocity variables, 68$\pm$11\% , is much larger
than the value expected on the basis of our simulations for a binary fraction of
60\%, as measured in the solar neighborhood. This indicates that the binary
fraction among the CEMP-s stars in the sample under consideration is likely to
be larger than that found among a randomly selected sample of metal-poor field
stars.

Another possibility to explain our finding is that the binary fraction amongst
CEMP-s stars is similar to that found for normal halo field stars, $\sim$60\%,
but that the orbital period distribution for CEMP-s stars in double systems is
different from that measured for binary field stars, peaking at much shorter
values. For a detected binary fraction of 68$\pm$11\%, this would
require a success rate in identifying binaries of $\geq$95\%. To achieve such a
high success rate with our observing pattern, the maximum period would have to
be of $\sim$6 years ($\log(P)=$3.4), as Table 4 shows. We cannot in principle 
rule out a maximum period this short, which might be consistent with the enrichment
scenario via wind-accretion (Han \etal{} 1995). However, the fact that 
a couple of the periods determined for CEMP-s are considerably longer than this value 
argues against it. Moreover, under this scenario
$\sim$40\% of the CEMP-s population must be non-binary, and no plausible
explanation exists for the chemical enrichment of the {\it s}-process elements
in these systems, which would appear completely analogous to that due to binaries.

When the input binary fraction is set to 100\%, the expected detection of
V$_{r}$ variables from our observational pattern rises to about 36\% of the
total. This arises since the observations collected so far for the sample under
analysis have a baseline of {\it at most} $\sim$12 years (and in many cases much
less), which is quite short, considering that the period distribution peaks at
$\log(P)=$4.8, {\it i.e.} $\sim$120 years. A considerable fraction of actual
binaries have periods that are too long to result in detectable V$_{r}$
variations using the available instruments over the time intervals explored.

\section{Orbital Period Limitations \label{s_bin_per}}

The observational result of 68$\pm$11\% radial-velocity variables obtained
from our sample exceeds the detection fraction even when all the stars in the
simulation are assumed to be binaries. This is a further argument in favor of a
binary scenario for the formation of CEMP-s stars. 

This result leads to speculation about the period distribution of the stars
under consideration. Most likely, {\it all} CEMP-s stars are in binary systems,
and owe their chemical peculiarities to the accretion of processed material from
a post-AGB evolved companion. If this scenario is correct, the semi-major axis
of their orbits (and therefore their orbital periods) must lie within the useful
range of values where such accretion processes are thought to take place. The
separation {\it must} exceed the stellar radius of the presumed donating
companion during its previous evolutionary phases. In fact, if the separation
were smaller than the RGB radius of the evolved companion, mass transfer would
take place during that phase and affect subsequent evolution, preventing the
donor star from undergoing its normal AGB phase.
This phenomenon indeed exists,
and is referred to as the McCrea transfer mechanism (McCrea 1964); its outcome
would likely be to convert the close pair of stars into a blue straggler. In
fact, as shown by Carney \etal{} (2001), field blue stragglers share similar
properties with Ba II, classical CH stars, and subgiant CH stars, {\it i.e.} they
are members of long-period, low-eccentricity binaries, suggesting that mass
transfer has been involved in their formation. Ryan \etal{} (2001) and Ryan \etal{} 
(2002) were led to a similar conclusion concerning field blue stragglers by
considering the depletion of Li during a mass-transfer episode, and spin-up of
the surviving star. However, in the present case it is necessary for the donor
star to pass through its AGB phase in order for the {\it s}-process elements to
be synthesized\footnote{C-enhanced, {\it s}-process enhanced blue stragglers are
known (Sneden, Preston, \& Cowan 2003); Luck \& Bond (1991) proposed that
some of them evolve into sgCH stars}.

On the other hand, the value of the orbital separation must be small enough to
allow for capture of a sufficient amount of processed material to create the
observed chemical enhancements in carbon and the $s$-process elements.\\ 
A
reasonable value for the lower limit can be set by adopting the RGB tip radius.
Using the Y$^2$ (Yi, Kim, \& Demarque 2003) database and an $\alpha$-enhanced
isochrone, [$\alpha$/Fe]=0.3 with [Fe/H]= $-$ 2.5, we obtain a value of
$\sim$0.5\,AU, which in turn leads to a limit on $\log(P)$ of $\sim -$0.65. The
upper limit to the useful interval requires detailed modeling of the enrichment
mechanism (see, {\it e.g.}, Han \etal{} 1995) and depends on the evolutionary
state of the accreting star. An approximate estimate for the most extreme value
predicted by Cristallo (2004) is $\log(P) \sim$5.4; however, given the
complexity of the assumptions involved, such a value must be considered as only
a very rough estimate.

Table \ref{t_bin_probp} shows the results of the same simulations described in
\S 4, but applying an orbital period cutoff.  In other words, the Duquennoy \& Mayor (1991)
orbital period distribution was adopted, but the permitted values of $\log(P)$
are limited to ranges whose lower cutoffs range between \lgpmin$=-$1.0 and 2.0,
while the upper ones are between \lgpmax$=$2.6 and 6.6. It is important to keep
in mind that the application of sharp cutoffs is only a very rough approximation. In
fact, a more accurate approach would require the convolution of the period
distribution function for binary field stars with theoretical Roche-lobe
overflow (for short periods) and wind (for long periods) accretion efficiencies,
which would reflect the C and {\it s}-process enhancements as a function of the
period. Thus the period distribution for these objects would taper off, rather
than truncate, at high values of $\log P$. However, we are not aware of the
existence of any systematic accretion efficiency calculation, therefore we adopt
simple cutoffs for our simulations, which, although not strictly accurate,
provides an interesting comparison with the observations.

According to our simulations, the fraction of binaries expected to be detected
with the observational schemes used to observe the sample stars, assuming the
period cutoffs discussed in the text ($-$0.65$<\log(P)<$5.4) along with the
Duquennoy \& Mayor (1991) distribution, is about 60\%, compatible with the
observed value. If the upper limit is shortened to 5.0 ($-$0.65$<\log(P)<$5.0), 
the expected binary detection fraction is 67\%, very similar to our 
observational finding. 
This result does indeed partially
depend on the fact that about half of the sample 
is made up of objects with derived orbital parameters and 
periods much shorter than 
typical values in the field population, as provided by the Duquennoy 
\& Mayor (1991) 
distribution. However, it should be noted that 
this is not a bias for the sample, as 
the objects were selected {\it only} on the basis of their chemical 
and evolutionary characteristics, with no previous knowledge
of their binary status.

It should be noted that the predicted detection fraction values are much
more sensitive to the adopted upper limit on the period than on the lower limit.
In fact, a change of a factor of 100 in the upper cutoff, bringing it from
$\log(P) = $6.2 to 4.2, would increase the expected detection rate by over
$40$\%, while moving the lower limit from 2.0 to 0.0 increases the fraction by
less than 10\%. This is not surprising, given that the adopted period distribution
tails off to very small values for low values of $\log (P)$, while
the upper cutoffs lie around the maximum of the distribution. Hence, the large
observed binary fraction reflects (and constrains) the period distribution of
CEMP-s stars primarily at the high end of the distribution of possible values.
It is noteworthy that models predict that one of the effects of 
mass transfer is that of lengthening the orbital period.
Therefore, it is expected that the original period distribution
({\it i.e.} before mass tranfer took place) for these objects was likely 
shifted toward shorter values.

\section{Formation scenario for CEMP-s stars}

Analysis of a well-defined sample of CEMP-s stars has led to the identification
of a binary fraction which exceeds that expected if the actual proportion of
binaries in the sample were consistent with the measured binary fraction for
field stars (Jahreiss \& Wielen 2000). Our extensive simulations show that, with
our observational patterns, we should identify only about $\sim$22\% ({\it i.e.}
$\sim$4 stars out of 19) as radial-velocity variables, while we find that 14
stars out of 19 of our sample, 68$\pm$11\%, exhibit clear V$_{r}$
variations. This value is larger than that expected even for the case in which the
binary fraction of the population is 100\%. This provides very strong
evidence that the binary fraction among CEMP-s stars is higher than
that found in the field, suggesting that in fact {\it all} of these objects are
members of binary systems. We conclude that CEMP-s stars are indeed the metal-poor
equivalents of the classical CH stars (McClure \& Woodsworth 1990). Thus, the
source of their chemical peculiarities is likely to be 
the accretion of material processed by
the now-evolved more massive companion, which, during or after its AGB phase
transfers mass, either via Roche-lobe overflow or wind, to the star we now
observe as a CEMP-s object.

The discrepancy between the number of binaries identified in our sample and the
expected numbers computed from simulations which assume a binary fraction of
100\%, and adopt the observed orbital period distribution of normal field stars
(Duquennoy \& Mayor 1991), may be of signficance. Adopting the period range
$\log(P)$ between $-$0.65 to 5.0, we find that the expected numbers of
identified binaries is much closer to that which is observed. While we make no
claim that the true orbital period limits can be obtained using this method, the
available data suggest that the measured period distribution of Duquennoy \&
Mayor (1991) is not appropriate for this class of stars; the true orbital
distribution is likely peaked at shorter periods. This result is consistent with
chemical enrichment via a mass-transfer scenario. The orbital separation needs
to be large enough to allow the donor star to undergo its AGB phase, but small
enough to allow the accretion to take place with sufficient efficiency to create
the observed abundance patterns. 

Long-term radial-velocity monitoring
will allow for a further test our results, possibly leading to the 
determination of orbital elements for a wider sample of CEMP-s stars.
With a sufficiently large sample, a statistical analysis 
of the orbital elements (see McClure 1983) 
could provide indications of the masses of the 
companions of the observed stars, and test whether they are consistent 
with those typical of white dwarfs.
 
\acknowledgments

We are very grateful to our referee, Bruce Carney, for his comments and
suggestions, which greatly improved this paper. The authors thank Norbert
Christlieb, Judy Cohen, Sergio Cristallo, John Norris, and Oscar Straniero for
useful discussion. S.L., R.G., and E.C. acknowledge partial support from the
MURST COFIN 2001 and MURST COFIN 2002. ST gratefully acknowledges partial
support from a UK Universities Overseas Research Student Award (ORS/2001031002)
for his large PhD programme, and a travel grant to the TNG provided by the
European Commission through the ``Access to Research Infrastructure Action'' of
the ``Improving Human Potential Programme'' awarded to the Instituto de
Astrof\'isica de Canarias. T.C.B. acknowledges partial funding for this work
from grants AST 00-98508, AST 00-98549, and 04-06784, as well as from grant PHY 02-16783,
Physics Frontier Centers/JINA: Joint Institute for Nuclear Astrophysics, awarded
by the US National Science Foundation.

\clearpage

\begin{deluxetable}{lrrrr}
\tablecaption{Atmospheric Parameters For Sample Stars 
\label{t_star_at}}
\tablehead{
\colhead{StarID}&\colhead{T$_{\rm eff}$ (K)}&\colhead{$\log g$}&\colhead{[Fe/H]}&\colhead{Source}}
\startdata 
   CS~22880-074&    5850& 3.8& $-$1.93& 1\\
   CS~22881-036&    6200& 4.0& $-$2.10& 4\\
   CS~22898-027&    6250& 3.7& $-$2.25& 1\\
   CS~22942-019&    5000& 2.6& $-$2.64& 1\\
   CS~22948-027&    4800& 1.8& $-$2.47& 6\\
   CS~22956-028&    7035& 4.5& $-$1.90& 3\\
   CS~29497-030&    7050& 4.2& $-$2.16& 3\\
   CS~29497-034&    4980& 2.1& $-$2.60& 6\\
   CS~29509-027&    7050& 4.2& $-$2.02& 3\\
  CS~29526-110&     6500& 3.1& $-$2.38& 1\\
   CS~30301-015&    5250& 1.8& $-$2.25& 1\\
     HD~196944&     5250& 1.7& $-$2.40& 2\\
     HD~198269&     4800& 1.3& $-$2.20& 2\\
     HD~201626&     5190& 2.3& $-$2.10& 2\\
     HD~224959&     5200& 1.9& $-$2.20& 2\\
  HE~0024-2523&     6625& 4.3& $-$2.72& 7\\
  HE~2148-1247&     6380& 3.9& $-$2.50& 5\\
     LP~625-44&     5500& 2.8& $-$2.71& 8\\
      LP~706-7&     6600& 3.8& $-$2.74& 8\\ 
\enddata
\tablerefs{(1) Aoki \etal~\cite{aok3}; (2) Van Eck {\it et al.} \cite{vaneck03};
(3) Sneden, Preston, \& Cowan (2003); (4) Preston \& Sneden \cite{ps01};
(5) Cohen \etal~\cite{coh03}; (6) Hill \etal~\cite{hill00};
(7) Lucatello \etal~(2003); (8) Aoki \etal~\cite{aok01}
}
\end{deluxetable}

\clearpage
\begin{deluxetable}{llrrrc}
\tablecaption{Observation Log and Measured Radial Velocities\label{t_obs_log}}
\tablehead{\colhead{Star ID}&\colhead{MJD}&\colhead{Exp.}&\colhead{Instr.}&\colhead{V$_{r}$}&
\colhead{$\sigma_{({\rm V}_{r})}$}\\
\colhead{} &\colhead{} &\colhead{Time}&\colhead{}&\colhead{\kms}&\colhead{\kms}}
\startdata

CS~22956-028 & 52470.29 &1800\,s& UVES       & 24.60 &    0.20 \\
CS~29497-034 & 52471.31 &2700\,s& UVES       &$-$52.10 &    0.40 \\
CS~22880-074 & 52391.72 & 1200\,s& UES        & 59.29&    0.14\\
             & 52419.65& 1800\,s& SARG& 58.83& 0.22\\
             & 52487.51& 2100\,s& UES & 58.71& 0.26\\
 CS~22898-027& 52151.49& 2700\,s& UES &$-$48.41& 0.11\\
             & 52417.67& 1200\,s& SARG& $-$49.54& 0.25\\
             & 52487.47& 1200\,s& UES& $-$48.78& 0.28\\
CS~29526-110 & 51804.40& 12600\,s& UCLES& 186.16& 0.19\\
             & 52152.73& 900\,s& UES& 201.83& 0.26\\
CS~30301-015 & 52152.35& 1200\,s& UES& 85.66& 0.12\\
             & 52390.55& 1800\,s& UES& 85.28& 0.14\\
HD~196944    & 52419.72& 300\,s& SARG& $-$169.29& 0.08\\
             & 52487.46& 300\,s& UES& $-$168.49& 0.11\\
LP~625-44    & 52150.35& 300\,s& UES& 28.06& 0.12\\
             & 52390.68& 450\,s& UES& 26.66& 0.10\\
             & 52417.74& 480\,s& SARG& 26.34& 0.30\\
             & 52487.39& 600\,s& UES& 27.48& 0.22\\
LP~706-7     & 52150.75& 300\,s& UES& 79.48& 0.15\\
             & 52487.70& 1200\,s& UES& 79.53& 0.17\\ \enddata
\end{deluxetable} 
\clearpage
\begin{deluxetable}{lrrrrrrrr} 
\tablecaption{Probability of Radial Velocity Variations \label{t_bin_po}}
\tablehead{ \colhead{Star}
&\colhead{Baseline}&\colhead{f}&\colhead{$\chi^{2}$}
&\colhead{P($\chi^{2}|f$)}&\colhead{Q($\chi^{2}|f$)}
&\colhead{Source}&\colhead{Orbital}&\colhead{Source}\\
 \colhead{ID}
&\colhead{(days)}&\colhead{}&\colhead{}
&\colhead{}&\colhead{}
&\colhead{}&\colhead{solution?}&\colhead{}} \startdata
   CS~22880-074&   3662&   17&    1.698&1.000 &    0.000&1, 2, 5 & No &\\
   CS~22881-036&   2561&   13&    1.645&1.000 &    0.006&6       & No &\\
   CS~22898-027&   4737&   15&    4.553&0.995 &    0.008&1, 2, 5 &No  &\\
   CS~22942-019&   3665&   15&  118.234&0.000 &    1.000&1, 5    &Yes &5 \\
   CS~22948-027&   2560&   23&   42.733&0.007 &    0.993&6, 10,13&Yes &5\\
   CS~22956-028&   3636&   24&  499.586&0.000 &    1.000&2, 4, 6 &Yes &4\\
   CS~29497-030&   3313&   16&   69.791&0.000 &    1.000&4, 6    &Yes &4\\
   CS~29497-034&   3020&    9&   48.512&0.000 &    1.000&2, 8, 14&Yes &14\\
   CS~29509-027&    351&    2&   40.914&0.000 &    1.000&4       &Yes &4\\
  CS~29526-110&     348&    2&  314.863&0.000 &    0.994&1, 2    &No  &\\
  CS~30301-015&     275&    5&    1.003&0.606 &    0.431&1, 2    &No  &\\
     HD~196944&     683&    4&   47.579&0.000 &    1.000&1, 2, 3 &No  &\\
     HD~198269&    2351&   17&   96.683&0.000 &    1.000&3, 12   &Yes &12\\
     HD~201626&    3352&   26&   85.147&0.000 &    1.000&3, 12   &Yes &12\\
     HD~224959&    2934&   15&  156.755&0.000 &    1.000&3, 12   &Yes &12\\
  HE~0024-2523&     399&   17& 4955.382&0.000 &    1.000&9       &Yes &9\\
  HE~2148-1247&     364&    3&    8.720&0.033 &    0.967&7       &No  &\\
     LP~625-44&    5183&   12& 1088.632&0.000 &    1.000&2, 10,11&No  &\\
      LP~706-7&    4433&    7&    8.428&0.296 &    0.704&2, 11   &No  &\\ 
\enddata
\tablerefs{(1) Aoki \etal~\cite{aok3}; 
(2) Present work; (3) Van Eck {\it et al} \cite{vaneck03};
(4) Sneden, Preston, \& Cowan (2003); (5) Preston \& Sneden \cite{ps01};
(6) Preston \& Sneden \cite{ps00}; (7) Cohen \etal~\cite{coh03}; (8) Hill \etal~\cite{hill00};
(9) Lucatello \etal~(2003); (10) Aoki \etal~\cite{aok01}; (11) Norris, Ryan, \& Beers (1997a);
(12) McClure \& Woodsworth (1990); (13) Aoki \etal{} 2002a; (14) Barbuy \etal{} 2004.}
\end{deluxetable}
\clearpage

\begin{deluxetable}{lrcrlrrc}
\tablecaption{Orbital Elements for Sample Stars\label{t_bin_orb}}
\centering
\tablehead{\colhead{Star ID} &\colhead{P(days)}&
\colhead{K$_1$(\kms)} &\colhead{$\omega$(deg)} &
\colhead{JD$_{0}$}&\colhead{V$_{\gamma}$(\kms)}&
\colhead{$e$}&\colhead{Source}}
\startdata
CS~22942-019 & 2800 & 5.0 & 280 & 2439390&$-$237.7& 0.10& 1\\ 
CS~22948-027 &  505 & 4.0 &  78 & 2448110&$-$66.2 & 0.30& 1\\
CS~22956-028 & 1290 & 8.5 &266  &2448831.0&   34.0 & 0.22& 2 \\
CS~29497-030 & 342  & 4.1 & 120 &2448500.0&   45.0 & 0.00& 2\\
CS~29497-034 &4130  & 5.2 & 13  &2449800  &$-$47.5 & 0.02& 3\\
CS~29509-027 & 194  & 3.8 & 194 &2448624.0&   74.2 & 0.15& 2\\
HD~189269    &1295  & 9.3 & 352 &2446358  & $-$203.39&0.094& 4\\
HD~201626    & 407  &12.1& \nodata&2445858.3&$-$378.77&0&4\\
HD~224959    &1273  &9.0 &10   &2446064  &$-$127.85&0.179&4\\
HE~0024-2523 & 3.41 &51.9 &\nodata&252059.596&$-$178.3&0&5\\
\enddata
\tablerefs{(1) Preston \& Sneden \cite{ps01}; (2)  Sneden, Preston, \& Cowan (2003);
(3) Barbuy \etal{} 2004; (4) McClure \& Woodsworth (1990); (5) Lucatello \etal~(2003)}
\end{deluxetable}
\clearpage

\begin{deluxetable}{llll}
\tablecaption{Expected Fraction of V$_{r}$ Variable stars
as a Function  of Adopted Binary Fraction\label{t_bin_proc}}
\centering
\tablehead{\colhead{Binary} &\colhead{Detection}&\colhead{Binary} 
&\colhead{Detection}\\
\colhead{fraction}&\colhead{fraction}&\colhead{fraction}&\colhead{fraction}}
\startdata
0.0& 0.001&0.6& 0.217\\ 
0.1& 0.036&0.7& 0.256\\
0.2& 0.070&0.8& 0.286\\
0.3& 0.108&0.9& 0.322\\
0.4& 0.144&1.0& 0.356\\
0.5& 0.176&   &      \\
\enddata
\end{deluxetable}
\clearpage
\begin{deluxetable}{lllllllll}
\tablecaption{Expected Fraction of V$_{r}$ Variable 
for Different Period Cutoffs (days)\tablenotemark{a}
\label{t_bin_probp}}
\tablehead{\colhead{}&\colhead{\lgpmin}&\colhead{\lgpmin}&
\colhead{\lgpmin}&\colhead{\lgpmin}&\colhead{\lgpmin}&
\colhead{\lgpmin}&\colhead{\lgpmin}\\
\colhead{\lgpmax}&\colhead{$-$1.0}&\colhead{$-$0.5}&
\colhead{0.0}&\colhead{0.5}&\colhead{1.0}&
\colhead{1.5}&\colhead{2.0}}
\startdata
 2.6&  0.970&   0.969&   0.967&  0.966&   0.963&   0.957&   0.948\\
 3.0&  0.965&   0.960&   0.959&  0.957&   0.954&   0.949&   0.945\\
 3.4&  0.950&   0.947&   0.949&  0.946&   0.941&   0.937&   0.929\\                
 3.8&  0.920&   0.919&   0.916&  0.914&   0.914&   0.906&   0.894\\
 4.2&  0.856&   0.855&   0.854&  0.851&   0.844&   0.830&   0.813\\
 4.6&  0.764&   0.763&   0.757&  0.756&   0.741&   0.726&   0.704\\
 5.0&  0.674&   0.672&   0.667&  0.662&   0.645&   0.626&   0.596\\
 5.4&  0.596&   0.593&   0.588&  0.579&   0.564&   0.543&   0.508\\
 5.8&  0.534&   0.528&   0.526&  0.518&   0.502&   0.478&   0.445\\
 6.2&  0.490&   0.448&   0.486&  0.478&   0.456&   0.439&   0.407\\
 6.6&  0.458&   0.455&   0.451&  0.440&   0.424&   0.404&   0.374\\
\enddata
\tablenotetext{a}{Assuming that all CEMP-s are binaries}
\end{deluxetable}
\end{document}